\documentstyle[preprint,aps]{revtex}

\begin{document}

\draft

\title{Stress-energy of a quantized scalar field in static wormhole
spacetimes}

\author{Brett E.\ Taylor\cite{me} and William A.\ Hiscock\cite{doc}}

\address{Department of Physics, Montana State University, Bozeman,
Montana 59717}

\author{Paul R.\ Anderson\cite{And}}

\address{Department of Physics, Wake Forest University, Winston-Salem,
North Carolina 27109}

\date{August 12, 1996}

\maketitle

\begin{abstract}

Static traversable wormhole solutions of the Einstein equations
require ``exotic'' matter which violates the weak energy condition.
The vacuum stress-energy of quantized fields has been proposed as the
source for this matter.  Using the Dewitt-Schwinger approximation,
analytic expressions for the stress-energy of a quantized massive
scalar field are calculated in five static spherically symmetric
Lorentzian wormhole spacetimes.  We find that in all cases, for both
minimally and conformally coupled scalar fields, the stress-energy
does not have the properties needed to support the wormhole geometry.

\end{abstract}


\section{Introduction}

The subject of Lorentzian wormholes has been of some interest in
recent years.  Since the publication of the general form for the
static spherically symmetric Lorentzian wormhole by Morris and
Thorne\cite{MT} a great deal of work has been done in an effort to
understand whether such solutions to the Einstein equations are
compatible with the known laws of physics.  While the very existence
of such solutions is clearly exotic, in allowing alternate
``shortcut'' paths connecting otherwise distant regions of the
universe, even more disturbing properties of Lorentzian wormholes have
been noted.

Following the initial publication of Morris and Thorne, a number of
groups\cite{MTY,Nov} showed that it was possible to utilize these
solutions to construct a spacetime with closed timelike curves(CTC)--a
``time machine''.  This discovery spurred a large amount of work on
whether and how one can make sense of various sorts of physical
processes\cite{Echev,Fried1,Fried2} that may take place in such
spacetimes.  Others\cite{Kim,Hawking,Visser} have attempted to
determine if the known laws of physics will prevent the conversion of
a initially chronal wormhole (or other) spacetime into one in which
closed timelike curves are present-- the ``chronology protection
conjecture''.  Excluding the ill-understood possibility of quantum
gravity acting as a chronology protection agent, it appears that the
only possible source of chronology protection would be due to
diverging vacuum energy of quantized fields on the chronology horizon
of spacetimes where CTCs are formed\cite{HisKon}, though the issue is
not resolved.  It is not clear whether the gravitational backreaction
to the divergences is sufficiently strong to alter the spacetime
causal structure, preserving chronology, before quantum gravity
becomes important.  It also appears that divergences may be avoided
for at least some quantized fields\cite{TanHis2}.

A fundamental difficulty associated with the wormhole solutions, even
in the absence of closed timelike curves, is the nature of the
stress-energy tensor associated with such a geometry.  Morris and
Thorne\cite{MT} demonstrated that the stress-energy tensor of a static
spherical wormhole must satisfy two conditions at the wormhole throat.
First, it was shown that the radial stress must be negative, {\it
i.e.}, a tension $\tau$ rather than a pressure; second, the tension
must be larger in magnitude than the value of the energy density at
the throat ($\tau > \rho$).  They termed matter obeying these two
conditions ``exotic'', since matter obeying the second condition will
necessarily violate the weak energy condition.  Such behavior is not
observed in known forms of classical matter; however, it has been
repeatedly speculated that the stress-energy of quantized fields might
satisfy these conditions, supporting traversible wormholes.

The purpose of this paper is to examine whether the stress-energy of
quantized fields in fact will have the appropriate form to support a
traversible wormhole geometry.  Calculating the stress-energy tensor
of a quantized field in a curved spacetime is a difficult and arduous
task.  In this paper, the DeWitt-Schwinger approximation will be used
to evaluate the stress-energy tensor of a quantized massive scalar
field in five candidate wormhole geometries.  Recent work by Anderson,
Hiscock, and Samuel\cite{big} has demonstrated that the
DeWitt-Schwinger approximation is quite accurate so long as the radius
of curvature of the spacetime is greater than the Compton wavelength
of the massive field. In contrast, the various analytic approximations
for massless fields are not as robust, and, in all but Ricci flat
spacetimes, contain an arbitrary parameter whose value cannot
be fixed except by experiment\cite{big}.

We find that in all five wormhole geometries, for the most physically
plausible values of the curvature coupling -- minimal ($\xi = 0$) or
conformal ($\xi=1/6$) coupling -- the stress-energy tensor
of the quantized massive scalar field is never of the correct form to
support the wormhole.  Three of the wormhole geometries contain
adjustable parameters in the metric; our conclusion holds for all
values of these parameters.  In each case, either the radial stress is
positive (pressure rather than tension) or the magnitude of the
tension is less than that of the energy density (not ``exotic''). 

If one allows arbitrary values of the curvature coupling, $\xi$, then
in three of the metrics examined it is possible to find cases where
the stress-energy does have the form to help support the wormhole. In
the other two cases, the stress-energy fails to support the wormhole
for all possible values of $\xi$.

It thus appears questionable whether the vacuum stress-energy of
quantized fields can in fact supply the exotic matter required for
a traversable wormhole. For physically plausible, minimally or 
conformally coupled fields in the cases examined, the vacuum 
stress-energy of a quantized massive field would actively oppose
the formation or maintenance of such a wormhole. Since vacuum
energies, unlike classical fields, cannot be ``engineered away'',
whatever form of matter present which would satisfy the exotic
conditions must also overcome the vacuum energy contributions of
fields such as these, to yield a total stress-energy for the
sum of all fields present which obeys the exotic conditions.

The general static spherically symmetric Lorentzian wormhole spacetime
is described in Sec.\ II.  The conditions which the stress-energy of
the quantized field must satisfy if it is to help hold the wormhole
open are developed there in terms of the parameters describing the
spacetime.  The calculation of the stress-energy for a quantized
scalar field in a general static spherically symmetric spacetime is
discussed as well as the range of validity for these approximation
methods.  In Sec.\ III the five separate wormhole spacetimes for which
the stress-energy tensor of the quantized scalar field was calculated
are described and the DeWitt-Schwinger expressions for the stress-
energy of quantized massive scalar fields for each spacetime are
given.  Throughout the paper we use units such that $\hbar = c = G =
1.$ Our sign conventions are those of Misner, Thorne, and Wheeler
\cite{MTW}.

\section{General wormhole spacetimes and calculations of stress-energy
for static spherically symmetric spacetimes}

In 1988, Morris and Thorne\cite{MT} published the general form for a
static spherically symmetric Lorentzian wormhole.  It is given by the
line element
\begin{equation}
	ds^2 = -e^{2\Phi(r)}dt^2 + \left({1 -{ b(r)\over r}}
        \right)^{-1}dr^2 + r^2d\theta^2 + r^2 sin^2\theta d\phi^2.
\label{line}
\end{equation}
There are two arbitrary functions in the line element:  $\Phi(r)$,
which is called the redshift function as it describes the
gravitational redshift in this spacetime; and $b(r)$, which is called
the shape function as it describes the shape of the spatial geometry
of the wormhole in an embedding diagram.  In addition the coordinate
$r$ is constrained to run between $r_0 \leq r < \infty$, where $r_0$
is the throat radius.  In a static orthonormal frame, the energy
density and radial tension are found to be
\begin{equation}
	\rho \equiv T_{\hat t \hat t} = {b' \over 8\pi r^2} ,
\label{rho}
\end{equation}
\begin{equation}
	\tau \equiv -T_{\hat r \hat r} = {1 \over 8\pi}
        \left[{{b \over r^3}- 2\left({1-{b \over r}}\right)
        {\Phi ' \over r} }\right],
\label{tens}
\end{equation}
where a prime represents a derivative with respect to $r$.  Morris and
Thorne also proved that the matter associated with the wormhole by the
Einstein equations must satisfy two conditions.  First, at the throat,
\begin{equation}
	\tau_0 > 0,
\label{cond1}
\end{equation}
the tension must be positive.  Second, the matter in the neighborhood
of the throat must be ``exotic'' in the sense that:
\begin{equation}
{\tau_0 - \rho_0 \over \vert \rho_0 \vert} > 0
\label{cond2}
\end{equation}
where the subscript indicates that the quantities are evaluated at the
throat.  Such matter inevitably must violate the weak energy
condition; that is, some timelike observers will measure local energy
densities to be negative.  Our primary goal in this paper is to
examine the stress-energy tensor of a quantized field in several
wormhole spacetimes, and to determine whether that stress-energy
satisfies Eqs.(\ref{cond1},\ref{cond2}).  If so, then the quantized
field is acting in such a way as to maintain the wormhole geometry.
If not, then the vacuum state stress-energy of the quantized field
would act in a fashion so as to close off the throat and (in a
self-consistent treatment) presumably hinder the formation of
macroscopic wormholes.

Methods for calculating the stress-energy tensor for quantized fields
in static spherically symmetric spacetimes have been developed by a
number of groups.  Approximate analytic expressions for $\langle
T_{\mu\nu} \rangle$ for conformally invariant massless scalar, spinor,
and vector fields in Einstein spacetimes (for which $R_{\mu\nu} =
\Lambda g_{\mu\nu}, \Lambda = constant)$ have been found by Page,
Brown, and Ottewill\cite{page,pbo}.  Frolov and Zel'nikov
\cite{frozel} have developed a geometrically based analytic
approximate expression for $\langle T_{\mu\nu} \rangle$ for
conformally invariant massless fields in static spacetimes.  The only
direct input to their approximation from quantum field theory is in
the form of the trace anomaly.  Direct, rather than approximate,
calculation of $\langle T_{\mu\nu}\rangle$ is generally a very
difficult task in curved spacetimes.  Howard and Candelas calculated
$\langle T_{\mu\nu} \rangle$ for a conformal massless scalar field in
the Schwarzschild spacetime\cite{howcan1,howcan2}.  Such direct
calculations are of course the only way the possible validity of the
analytic approximations can be firmly established.

Recently Anderson, Hiscock, and Samuel\cite{little,big} have
developed a method and numerical program able to calculate the
stress-energy tensor of a quantized scalar field with arbitrary
curvature coupling and mass in a general static spherically symmetric
spacetime.  In the course of developing this method, an analytic
approximation schemes for calculating the stress-energy tensor for a
massive scalar field with arbitrary curvature coupling was developed. 

The approximation developed amounts to an alternate derivation of the
DeWitt-Schwinger approximation utilizing a WKB expansion.  In a static
spherically symmetric spacetime, the large $m$ limit of the WKB
approximation for $\langle T_{\mu}$$^{\nu}\rangle$ for the massive
scalar field is equivalent to the DeWitt-Schwinger expansion.  In
order to obtain the expansion to order $1/m^2$ it is necessary to
carry the WKB expansion out to sixth order.  This approximation,
denoted by $\langle T_{\mu}$$^{\nu}\rangle_{DS}$ was found to be
exceedingly accurate in Reissner-Nordstr\"om black hole spacetimes as
long as the product of the mass of the field, $m$, and the mass of the
black hole, $M$, was greater than unity.  For example, choosing $Mm
=2$ gave fractional accuracy in the values of $\langle T^{\mu}$$_{\nu}
\rangle_{DS}$ of order $10^{-2}$ everywhere outside the event horizon.
The approximation is described in detail in Ref.\cite{big}.

\section{Stress-energy for quantized scalar fields in wormhole
spacetimes}

In this section analytic approximate expressions for the stress-
energy tensor of a quantized massive scalar field are evaluated
using the DeWitt-Schwinger method\cite{big} in five different
exemplar wormhole spacetimes.  The DeWitt-Schwinger approximation
is carried out to order $1/m^2$.  
                                                
The five different spacetimes are characterized by particular
choices of redshift function $\Phi(r)$, and shape function $b(r)$.
The energy density and radial tension are defined as:

\begin{equation}
\rho = - \langle T^{t} _{t} \rangle
\label{defrho}
\end{equation}

\begin{equation}
\tau = - \langle T^{r} _{r} \rangle.
\label{deftau}
\end{equation}

\subsection{Zero tidal force Schwarzschild wormhole}

A particularly simple set of wormhole geometries are those for which
\begin{equation}
	\Phi(r) = 0 .
	\label{ztfschphi}
\end{equation}
For these cases all stationary observers experience zero tidal forces.
A simple particular example is where the spatial geometry is chosen to
have the Schwarzschild form:
\begin{equation}
	b(r) = constant = r_0 ,
	\label{ztfschb}
\end{equation}
where $r_0$ is the throat radius.  Since $b$ is constant in this
solution, the background energy density $\rho$ vanishes identically,
by Eq.(\ref{rho})

The DeWitt-Schwinger approximation provides the following expressions
for the stress-energy tensor components of a quantized massive scalar
field in this spacetime:

\begin{equation}
	\langle T^{t} _{t} \rangle  = {r_0^2 \left(-405r + 448r_0 
	+2520r\xi - 2772r_0 \xi \right) \over 53760\pi^2r^{9}m^2} ,
	\label{masstconst}
\end{equation}
\begin{equation}
	\langle T^{r} _{r} \rangle  = {r_0^2 \left(261r - 238r_0 
	- 1008r\xi + 924r_0\xi\right) \over  53760\pi^2r^{9}m^2} ,
	\label{massrconst}
\end{equation}
\begin{equation}
	\langle T^{\theta} _{\theta} \rangle  = {r_0^2 \left(-783r 
	+ 833r_0 + 3024r\xi - 3234 r_0\xi \right) \over  
	53760\pi^2r^{9}m^2} .
	\label{massthconst}
\end{equation}
From these equations it can be seen that the radial tension is
positive only if $\xi > 23/84$, while the ``exotic'' condition of
Eq.(\ref{cond2}) is satisfied only if $\xi < 10/84 $.  It is thus not
possible to simultaneously satisfy the two conditions of
Eqs.(\ref{cond1},\ref{cond2}) for any value of the curvature coupling.
Therefore, the stress-energy of the quantized field will never have the
form required to act in support of the wormhole in this case.

\subsection{The simple wormhole}

This wormhole metric was used on a final examination in an
introductory relativity class at Caltech; it is discussed in Box 2 and
on p.  400 of Ref.  \cite{MT}.  The metric functions are
\begin{equation}
	\Phi(r) = 0 ,
	\label{simplephi}
\end{equation}
\begin{equation}
	 b(r) = {r_0^2 \over r} .
	\label{simpleb}
\end{equation}
We find the stress-energy tensor components for the massive scalar
field to be:
\begin{eqnarray}
	\langle T^{t} _{t} \rangle & = &{{r_0^2} \over {20160\pi^2 
	r^{12}m^2}}\left[5940r^4-22932r^2r_0^2 + 18025r_0^4 
	\right. \nonumber \\
	& &{} \left.
	+ \xi\left(- 60480r^4 + 238168 r^2 r_0^2 
	-188930r_0^4\right)
	\right. \nonumber \\ 
	& &{} \left.
	+ \xi^2 \left(151200r^4 -631680r^2r_0^2 
	+ 513660r_0^4 \right)
	\right. \nonumber \\ 
	& & {} \left.  + \xi^3\left(141120r^2 r_0^2	
	-160440r_0^4 \right) \right],
	\label{masstro2}
\end{eqnarray}
\begin{eqnarray}
	\langle T^{r} _{r} \rangle & = & \frac{r_0^2}{20160\pi^2 
	r^{12}m^2}\left[-2484r^4 + 7116r^2r_0^2 - 4445r_0^4 
	\right. \nonumber \\ &&  {} \left.
	+ \xi \left(24192r^4 - 69048r^2r_0^2 + 43050r_0^4\right)
	\right. \nonumber \\ &&  {} \left.
	+ \xi^2 \left(-60480r^4 + 181440r^2r_0^2 - 115500r_0^4
	\right)	\right. \nonumber \\ &&  {} \left.
	+ \xi^3 \left(- 40320r^2r_0^2 + 36120r_0^4\right) 
	\right],
 	\label{massrr02}
\end{eqnarray}
\begin{eqnarray}
	\langle T^{\theta} _{\theta} \rangle& = &\frac{r_0^2}{20160
	\pi^2 r^{12}m^2} 
	\left[7452r^4 - 28464r^2r_0^2 + 22225r_0^4 
	\right. \nonumber \\ &&  {} \left.
	+ \xi \left(- 72576r^4 + 276192r^2r_0^2 - 215250r_0^4
	\right) \right. \nonumber \\ &&  {} \left.
	+ \xi^2 \left(181440r^4 - 725760r^2r_0^2 + 577500r_0^4 
	\right)	\right. \nonumber \\ &&  {} \left.
	+\xi^3 \left(161280r^2r_0^2 - 180600r_0^4 \right) 
	\right]. 
	\label{massthr02}
\end{eqnarray}
In this case, the radial tension of the quantized massive scalar field
is positive at the throat if $\xi > 0.860358$; the exotic condition is
satisfied if either $\xi < 0.151551$ or $0.2596 < \xi < 1.42218$.
Both conditions are satisfied, and the wormhole is supported by the
stress-energy of the quantized field, only if $1.42218 > \xi >
0.860358$.  The stress-energy of a quantized massive field with either
conformal or minimal coupling does not have the form necessary to
support the wormhole throat.

\subsection{The ``absurdly benign'' wormhole}

In this case, 
\begin{equation}
	\Phi(r) = 0 ,
	\label{abphi}
\end{equation}
\begin{equation}
        b(r) = \frac{r_0\left(a + r_0 - r\right)^2}{a^2}, 
	\label{abb}
\end{equation}
where $r_0$ is again the throat radius and $a$ is an adjustable
length.  This spacetime is called ``absurdly benign'' by Morris and
Thorne, because classically all of the exotic material is contained in
the region $r_0 \leq r < r_0+a$.  The value of $b(r)$ given is valid
only within this range of values  of $r$.  For $r \geq r_0 + a$, the
shape function $b(r) = 0$ so outside this radius the spacetime is just
Minkowski space.

Due to the length of the algebraic expressions for the general
components of the stress-energy tensor, we will give only the values
of the components calculated at the throat, $r = r_0$, for the
remaining three spacetimes beginning with this spacetime.

The components of the stress-energy tensor for the massive scalar
field at the throat are found to be:

\begin{eqnarray}
	\langle T^{t} _{t} \rangle _0 & = & \frac{1}{161280\pi^2 
	a^5r_0^6m^2} \left[129a^5 + 1202a^4 r_0 + 5134a^3
	r_0^2 + 7856a^2 r_0^3 + 3884a r_0^4 + 264r_0^5 
	\right. \nonumber \\ & & {} \left.
	+ \xi \left(-756a^5 -11536a^4 r_0 - 59080a^3 r_0^2
	- 95200a^2 r_0^3 - 46368a r_0^4 - 2688 r_0^5\right) 
        \right. \nonumber \\ & & {} \left.	
	+\xi^2\left(23520a^4 r_0 + 191520a^3r_0^2 
	+ 342720a^2 r_0^3 + 164640a r_0^4  
	+ 6720r_0^5\right) 
	\right. \nonumber \\ & & {} \left.
	+\xi^3 \left(-161280a^3 r_0^2 - 349440a^2 r_0^3  
	- 161280a r_0^4 \right)\right] ,
	\label{masstabs}
\end{eqnarray}
\begin{eqnarray}
	\langle T^{r} _{r} \rangle _0 & = & \frac{1}{161280\pi^2 
	a^4 r_0^6m^2}\left[69a^4 + 450a^3 r_0 + 1878a^2 
	r_0^2 + 2272 a r_0^3 + 552r_0^4 
	\right. \nonumber \\ & & {} \left.
	+\xi\left(-252a^4 - 3360a^3 r_0 - 19824a^2 r_0^2 
	- 25536a r_0^3 - 5376r_0^4\right) 
	\right. \nonumber \\ & & {} \left.
	+\xi^2\left(6720a^3 r_0 + 70560 a^2 r_0^2 
	+ 94080a r_0^3 + 13440 r_0^4\right) 
	\right. \nonumber \\ & & {} \left.
	+ \xi^3\left(- 80640a^2 r_0^2 - 107520a r_0^3 \right)
	\right],
	\label{massrabs}
\end{eqnarray}
\begin{eqnarray}
	\langle T^{\theta} _{\theta} \rangle _0 &=& \frac{1}
	{80640\pi^2 a^5 r_0^6 m^2}\left[75a^5 + 876a^4 r_0 
	+ 3198a^3 r_0^2 + 5132a^2 r_0^3 + 2325a r_0^4
	+ 138r_0^5 
	\right. \nonumber \\ & & {} \left.
	+\xi\left(-315a^5 - 5838a^4 r_0 - 29484a^3 r_0^2 
	- 55776a^2 r_0^3 - 25200a r_0^4 - 1344r_0^5 \right)
	\right. \nonumber \\ & & {} \left.
	+\xi^2\left(10080a^4 r_0 + 84840a^3 r_0^2 + 198240a^2 
	r_0^3 + 85680a r_0^4 + 3360r_0^5\right) 
	\right. \nonumber \\ & & {} \left.
	+\xi^3\left(- 60480a^3 r_0^2 - 215040a^2 r_0^3 
	- 80640a r_0^4 \right)\right],
	\label{massthabs}
\end{eqnarray}
where the subscript zero denotes the value of a quantity at the 
throat.

The range of values of the curvature coupling constant which will
satisfy the conditions specified by Eqs.\ (\ref{cond1},\ref{cond2})
will depend on the value chosen for $a$. The regions in the $(\xi,a)$
plane for which the stress-energy conditions necessary for wormhole
support are satisfied are illustrated in Figure (1). For no value of
$a$ in this geometry will the tension be positive for either the 
minimally coupled or conformally coupled massive field.  Thus, 
a massive minimally or conformally coupled field never contributes
to the total stress-energy in a fashion so as to support the wormhole.

\subsection{Wormhole with finite radial cutoff in background
$T_{\mu\nu}$}

In this case a zero-tidal-force throat solution is joined at a finite
radius to an exterior Schwarzschild solution.  Since we are only
concerned with the stress-energy of the quantized fields in the
neighborhood of the throat, and our analytic approximate expressions
for $\langle T_{\mu\nu} \rangle$ are sufficiently local, only the
interior geometry in the neighborhood of the throat is needed.  There
\begin{equation}
	\Phi(r) = 0 ,
	\label{frcphi}
\end{equation}
\begin{equation}
	b(r) = r_0 \left(\frac{r}{r_0}\right)^{1-\eta} ,  
	\label{frcb}
\end{equation}
where $\eta$ is a constant bounded by $0 < \eta < 1$.

For the massive scalar field, the values for the stress-energy
components at the throat are given by:
\begin{eqnarray}
	\langle T^{t} _{t} \rangle _0 & = & \frac{1}{161280\pi^2
	 m^2 r_0^6} \left[64 + 112\eta - 490\eta^2 - 92\eta^3 
	 + 403\eta^4 + 132\eta^5 \right. \nonumber \\ & & {} 
	 \left.
	+\xi\left(- 672 - 1792\eta + 6860\eta^2 + 1148\eta^3  
	- 4956\eta^4 - 1344\eta^5 \right)
	\right. \nonumber \\ & & {} \left.
	+\xi^2\left(3360 + 10080\eta - 30240\eta^2 - 5040\eta^3 
	+ 18480\eta^4 + 3360\eta^5\right) 
	\right. \nonumber \\ & & {} \left.
	+\xi^3\left(-6720 - 20160\eta + 40320\eta^2 + 6720\eta^3  
	- 20160\eta^4\right)\right],
	\label{masstscal}
\end{eqnarray}
\begin{eqnarray}
	\langle T^{r} _{r} \rangle _0 & = & \frac{1}{161280\pi^2 
	m^2 r_0^6} \left[64 -210\eta^2 + 146\eta^3 + 69\eta^4 
	\right. \nonumber \\ & & {} \left.
	+\xi\left(-672 + 2940\eta^2 - 1848\eta^3 - 672\eta^4
	\right)	\right. \nonumber \\ & & {} \left.
	+\xi^2\left(3360 - 13440\eta^2 + 8400\eta^3 + 1680\eta^4
	\right)	\right. \nonumber \\ & & {} \left.
	+\xi^3\left(6720 + 20160\eta^2 - 13440\eta^3 \right)
	\right],
	\label{massrscal}
\end{eqnarray}
\begin{eqnarray}
	\langle T^{\theta} _{\theta} \rangle _0 & = & \frac{1}
	{161280\pi^2 m^2 r_0^6} \left[-128 + 504\eta - 840\eta^2 
	- 19\eta^3 + 495\eta^4 + 138\eta^5 
	\right. \nonumber \\ & & {} \left.
	+\xi\left( 1344 - 7392\eta + 11760\eta^2 + 462\eta^3 
	- 5460\eta^4 - 1344\eta^5 \right) 
	\right. \nonumber \\ & & {} \left.
	+\xi^2\left(- 6720 + 38640\eta - 53760\eta^2 
	- 840\eta^3 + 19320\eta^4 + 3360\eta^5\right)
	\right. \nonumber \\ & & {} \left.
	+ \xi^3\left(13440 - 70560\eta + 80640\eta^2 
	- 3360\eta^3 -20160\eta^4 \right)\right].
	\label{massthscal}
\end{eqnarray}
Despite having two adjustable parameters, $\eta$ and $\xi$, there
are no cases in which this stress-energy will satisfy both
conditions, Eqs.\ (\ref{cond1},\ref{cond2}). It thus appears that
the vacuum stress-energy of a massive scalar field will always
oppose this sort of wormhole.

\subsection {The Proximal Schwarzschild Wormhole}

This metric is similar to the Schwarzschild metric except for  
an additional term in $g_{tt}$, 
\begin{equation}
	-g_{tt} = 1 - \frac{r_0}{r} + \frac{\epsilon}{r^2} ,
	\label{PSgtt}
\end{equation}
\begin{equation}
	b(r) = r_0 .
	\label{PSb}
\end{equation}
The variable $\epsilon$ is a small positive constant.  The addition of
this term to the metric prevents the appearance of an event horizon in
this spacetime, keeping the wormhole traversable by Morris and
Thorne's definition\cite{MT}.  Due to its similarity to Schwarzschild,
this spacetime is called proximal Schwarzschild.
This metric is discussed in Ref. \cite{Visser2} in Sec. 13.4.3.

For a massive scalar field, the stress-energy components at the throat
are:

\begin{eqnarray}
	\langle T^{t} _{t} \rangle _0 & = & \frac{1}
	{322560\pi^2\epsilon^3 r_0^6 m^2} \left[-442\epsilon^3 
	- 491\epsilon^2 r_0^2 - 1044\epsilon r_0^4 + 358r_0^6 
	\right. \nonumber \\ & & {} \left.
	+ \xi\left(5600\epsilon^3 + 3332\epsilon^2 r_0^2 
	+ 10332\epsilon r_0^4  - 3402r_0^6\right)
	\right. \nonumber \\ & & {} \left.
	+\xi^2\left(- 14280\epsilon^3  - 4620\epsilon^2 r_0^2  
	- 31500\epsilon r_0^4 + 9870 r_0^6 \right)
	\right. \nonumber \\ & & {} \left.
	+\xi^3\left(- 26880\epsilon^3 - 10080\epsilon^2 r_0^2 
	+ 22680\epsilon r_0^4 - 5460r_0^6 \right)\right],
	\label{masstprox}
\end{eqnarray}
\begin{eqnarray}
	\langle T^{r} _{r} \rangle _0 & = & \frac{1}
	{322560\pi^2 \epsilon^3 r_0^6 m^2} \left[58\epsilon^3 
	- 107\epsilon^2 r_0^2 - 20\epsilon r_0^4 + 22r_0^6
	\right. \nonumber \\ & & {} \left.
	+\xi\left(112\epsilon^3 + 1428\epsilon^2 r_0^2 + 56\epsilon 
	r_0^4 - 210r_0^6\right) 
	\right. \nonumber \\ & & {} \left.
	+\xi^2\left( 840\epsilon^3 - 5460\epsilon^2 r_0^2 
	+ 630r_0^6\right) \right. \nonumber \\ & & {} \left.
	+\xi^3\left(- 6720\epsilon^3 + 5040\epsilon^2 r_0^2  
	- 420r_0^6 \right)\right],
	\label{massrprox}
\end{eqnarray}
\begin{eqnarray}
	\langle T^{\theta} _{\theta} \rangle _0 & = & \frac{1}
	{322560\pi^2 \epsilon^3 r_0^6 m^2} \left(-538\epsilon^3 
	- 382\epsilon^2 r_0^2- 833\epsilon r_0^4 + 283r_0^6
	\right. \nonumber \\ & & {} \left.
	+\xi\left(7112\epsilon^3 + 3220\epsilon^2 r_0^2 
	+ 9296\epsilon r_0^4 -3087r_0^6\right)
	\right. \nonumber \\ & & {} \left.
	+\xi^2\left(- 14280\epsilon^3 - 4620\epsilon^2 r_0^2  
	- 29400\epsilon r_0^4 + 9135r_0^6 \right) 
	\right. \nonumber \\ & & {} \left.
	+\xi^3\left(- 26880\epsilon^3  - 7560\epsilon^2 r_0^2  
	+ 20160\epsilon r_0^4 - 4830r_0^6 \right)\right].
	\label{massthprox}
\end{eqnarray}
If one assumes either minimal or conformal values for the curvature
coupling, $\xi$, then it is easy to show that there is no value of
$\epsilon$ which will result in a positive tension and also satisfy
the exotic condition of Eq.(\ref{cond2}).  Therefore, the vacuum
stress-energy tensor of any conformal or minimally coupled massive
scalar field will not help support the wormhole.  There are three
small regions in the $(\xi,\epsilon)$ plane in which the stress-energy
conditions necessary to support the wormhole will be satisfied.  These
regions are illustrated in Figure (2).  Therefore, the vacuum
stress-energy tensor of any conformal or minimally coupled massive
scalar field will not help support the wormhole.

\section{Discussion} 

We have shown that, within the context of the DeWitt-Schwinger
approximation for the vacuum stress-energy tensor of a quantized
massive scalar field, such a field will never have the needed
``exotic'' properties to support a static wormhole in the five
exemplar cases examined, if the field is either minimally or
conformally coupled to the scalar curvature.  In several of the cases
examined, the stress-energy tensor will not satisfy the exotic
conditions for any value of the curvature coupling, while for other
cases there are ranges of metric parameters and curvature couplings
which will result in a stress-energy tensor which is exotic in the
sense of Morris and Thorne.  While there is no experimental evidence
concerning values of the curvature coupling (indeed, there is a
general lack of evidence at the present for fundamental scalar
fields), other theoretical arguments have been made to demonstrate
that only small values of the curvature coupling, near zero, are
physically plausible.  For example, by requiring the contribution of a
quantized massless scalar field to the entropy of an equilibrium black
hole to be positive, it is possible to limit the curvature coupling to
the range $ -3.431 < \xi < 0.84$\cite{winnipeg}.

Our results are dependent upon the use of the DeWitt-Schwinger
approximation. This approximation has been shown to
be quite accurate and robust by direct comparison with exact
numerically calculated stress-energy tensors in black hole
spacetimes\cite{big}. So long as the mass of the scalar field
is large compared with the local radius of curvature of the
spacetime, we expect the analytic approximation to be very
close to the exact values which could be obtained numerically
(with great effort). In none of the cases examined does it
appear likely that the slight changes in going to exact
values for $\langle {T_\mu}^\nu \rangle$ would cause the
scalar field to satisfy the exotic conditions and support
the wormhole geometry.

In recent work, Ford and Roman\cite{ford} have used a bound they had
previously developed\cite{fordold} on negative energy densities in
Minkowski space to argue that a traversable wormhole must either be
microscopic--of order the Planck size--or the wormhole must have two
very different length scales, those being the size of the region where
the exotic matter is located and the throat radius.  They show that
for the absurdly benign wormhole that if one chooses a throat radius
of about two meters, one finds that $a$ must be about $10^{14}$
Planck lengths $\approx 10^{-19} cm$.  Using these values in our
results for the absurdly benign case yields interesting results.  For
the massive scalar field, the field acts to support the geometry only
if the curvature coupling is assigned obviously unphysical values,
namely $\xi > 1.25 \times 10^{20}$. 

Finally, while there may be other sorts of fields which will yield
vacuum stress-energies which satisfy the exotic conditions and thus
could aid in supporting a wormhole, it is important to note that 
the vacuum stress-energy of the fields studied here would also be
present in such a case, opposing the action of the hypothetical
wormhole-supporting fields. Thus, the present results not only 
indicate that massive quantized fields are unlikely to support
wormhole geometries, but in addition show that the vacuum 
stress-energy of such fields will oppose the formation and
maintenance of traversible wormholes.

\acknowledgements B.\ E.\ T.  would like to acknowledge helpful
discussions with Rhett Herman concerning the approximation method and
Tsunefumi Tanaka concerning general wormhole spacetime properties.
W.\ A.\ H.  would like to acknowledge help by John W.  Hiscock in
generating the data for Figures (1-2).  The work of W.\ A.\ H.  and
P.\ R.\ A.  was supported in part by National Science Foundation
Grants Nos.  PHY-9511794 and PHY-9512686, respectively.

\begin{figure} 
\caption{ Points within the shaded region represent values of $\xi$ and
$a$ for which the tension is positive and the exotic condition is
satisfied, so that the vacuum stress-energy is of the correct form to
help support the absurdly benign wormhole.}
\end{figure}

\begin{figure} 
\caption{ Points within the shaded regions represent values of $\xi$ and
$\epsilon$ for which the tension is positive and the exotic condition is
satisfied, so that the vacuum stress-energy is of the correct form to
help support the proximal Schwarzschild wormhole.}
\end{figure}

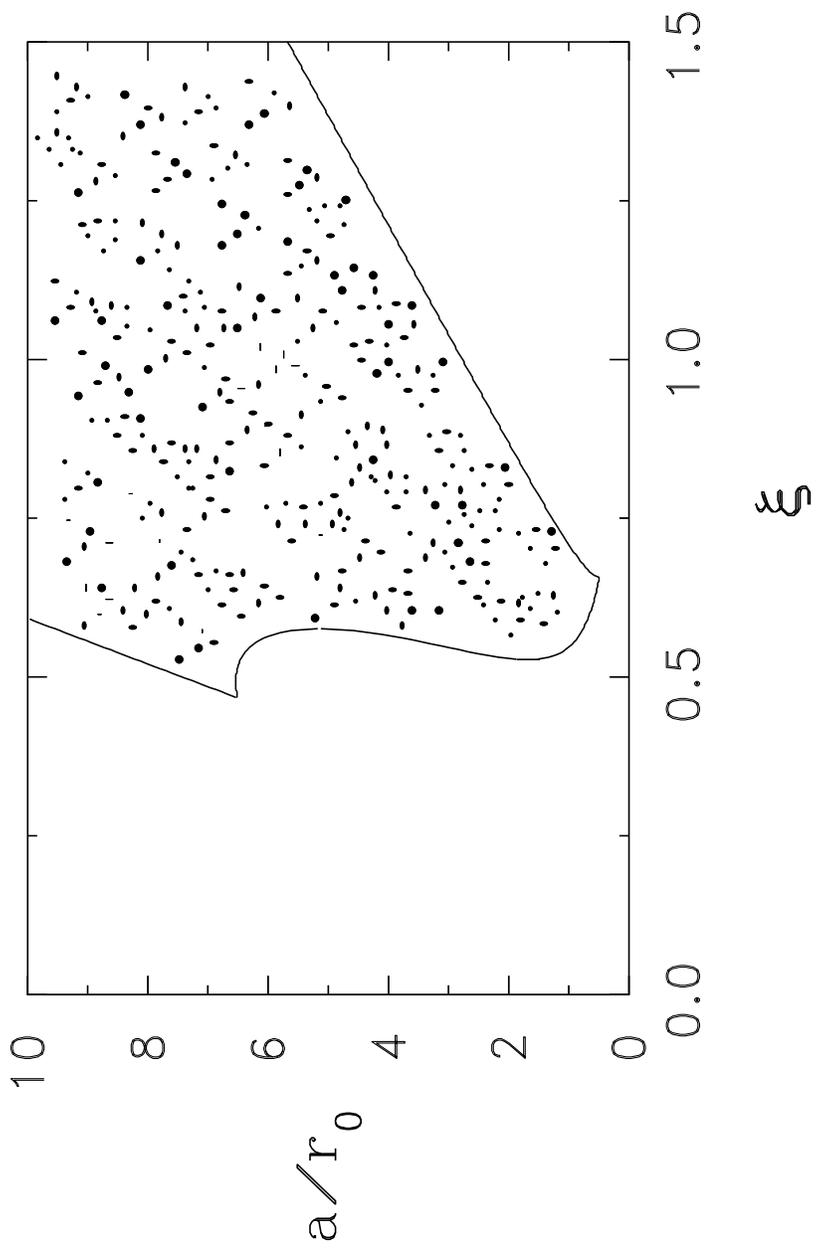
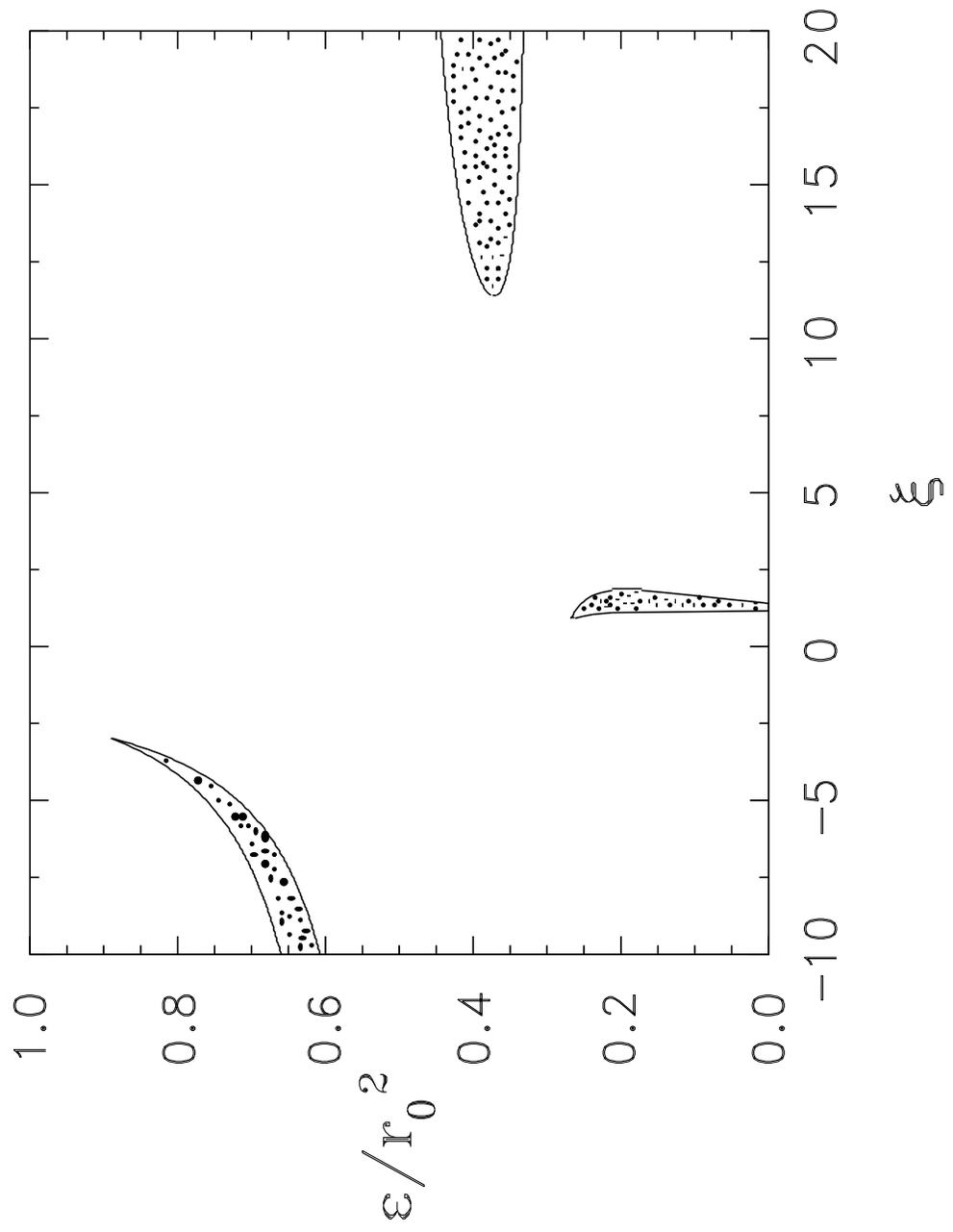
\end{document}